\documentstyle[12pt,aasms]{article}
\slugcomment{Submitted to Astrophysical Journal Letters}
\begin{document}
\title{An Anomalous UV Extension in NGC6251
\footnote{Based on
observations with the NASA/ESA Hubble Space Telescope
which is operated by AURA, Inc.,
under NASA contract NAS 5-26555}}

\author{P.~Crane and J.~Vernet}
\affil{ European Southern Observatory, Karl-Schwarzschild Strasse 2, D-85748 Garching,
Germany.}

\begin{abstract}
Deep U-band FOC images of the nuclear region of NGC6251 have revealed a region
of extended emission which is  most probably radiation scattered from a 
continuum source  in the nucleus. 
This radiation lies interior to a dust ring, 
is  nearly perpendicular to the radio jet axis, and is 
seen primarily  in the FOC U and b filters. The extension has a low observed
polarization($\le 10\%$), and is unlikely to
arise from line emission. We know of no other examples similar to what
we have found in NGC 6251, and we offer some tentative  explanations.
The nuclear morphology shows clear  similarities to 
that seen in the nucleus of NGC 4261 except for the extended U-band radiation.
\end{abstract}
\keywords{polarization; radiation mechanisms: non-thermal;\\
   galaxies: individual (NGC6251); galaxies: jets; ultraviolet: galaxies }

\section{Introduction}
Until the HST data were available, the  optical images of  NGC 6251 showed a
rather  normal elliptical galaxy although the presence of dust in the nuclear 
regions had been reported(\cite{nieto83}). This dust was shown to be confined 
to a ring or disk like structure by HST(\cite{oneil94}). Thus NGC6251 was 
and is thought to be merely another
case in which the dust features delineate the morphology and possibly the
dynamics of the nuclear regions of a galaxy which is otherwise unremarkable in
the optical. 

NGC6251 is classified as E2, lies at 
a distance of 114 Mpc (v= 7400 km/s $H_o = 65~km~s^{-1}~Mpc^{-1}$) and has an 
absolute magnitude of $M_v = -21.3$. Optical spectra(\cite{shuder81}
;~\cite{ANT84}) show low level
Seyfert activity if any at all.  The Seyfert classification has probably been 
prejudiced by the rather spectacular appearance of this galaxy in the radio.
Indeed, the radio images(\cite{per84}) reveal a highly collimated jet that 
extends to at least 1 Mpc and shows structure on every scale that 
it has been observed(\cite{jones94}). 
The X-ray data(\cite{baw93}) show evidence for an
unresolved($\leq 3''$) non-thermal nuclear component in the core. Optical 
emission associated with one of the bright radio knots has been detected about 
20 arcsecs from the nucleus (\cite{keel88}).

NGC 6251 was observed with the FOC on HST as part of a survey of ``normal'' 
elliptical galaxies(\cite{pceta93}). Important for the results
reported here, this survey  is so-far the 
only one with HST which includes two colors one of which is in the near UV.
The initial images were taken before the first repair mission and suffered 
severely from the spherical aberration. Nevertheless, it was noticed that
the UV(F342W) image and only this image showed an unexpected extension 
that did not seem to be 
an artifact related to the point spread function(\cite{pc93}). 
Subsequent pre-Costar
images also revealed the same extension(\cite{pc96}). Motivated by this 
anomaly, a set of post-Costar images was obtained that clearly revealed
the remarkable U-band emission inside the dust ring in NGC 6251 that is
reported here.

\section{Data and Analysis}

The data come  from our own FOC observations and from 
PC2 images available in the HST archive. Table 1 summarizes the observational
material.
Figure~\ref{comps} shows a composite of the FOC and PC2 images and clearly 
demonstrates the main discovery reported here; the emission extending to the
south of the nucleus and
which is seen so prominently in the F342W image. The presence of the
U-band emission is particularly remarkable considering its apparent
absence in the PC2 V and I images and it's  weakness in the F410M image. 
The anomalous  emission
is confined to the region interior to the dust ring. It extends about 0.5 
arcsec or 280 pc from the nucleus. The slight extension seen to
the North of the nucleus appears to be cut-off by the dust ring. This 
is particularly clear when the U  and V images are compared in detail
(seeFig.~\ref{dust}).  
Thus we suggest that the emission fills the inner regions of the dust
ring and is consequently not intrinsically asymetric;  the details of
what we observe are determined by the orientation and morphology of the
dust ring. To be precise,
the dust ring does not lie in a plane, but is bent on the NE side of the 
nucleus, and thus obscures the emission more effectively.
as can be seen in Fig.~\ref{dust}.
However, we cannot 
cannot rule out some clumping or patchy obscuration in the emission.

The FOC images contain interesting details of structure within 
about 20--40  pc of the nucleus. 
Also, there is some evidence for radiation emerging close to the 
radio axis, but this is very faint. We will 
discuss these in a subsequent communication(\cite{jv97a}). 

After the clear detection  of the extended U-band emission in the nucleus, 
FOC polarization
images were obtained since one potential explanation for the extended
emission was scattering 
of photons from a hot central source.  The polarization  data were
processed following the same procedures used for the Pictor A 
data(\cite{thoms95}). Figure~\ref{poldata} shows the 
resulting polarization map.
The low polarization in the extension
and the apparent lack of change along the extension imply 
that, if the emission is due to scattered photons, the scattering process must 
somehow produce a low  polarization. We put an
upper limit of $\leq 10\%$ on the polarization in the extended 
emission. By analyzing the 
the fractional polarization
statistical distribution in the SW extension,
we determine an average  polarization of  about 5\% but with
a big uncertainty.  We note 
that regions closer to the nucleus do show enhanced polarization as would
be expected if the radiation is scattered. We also note a small region
of enhanced brightness and polarization to the east of the nucleus in
the direction of the radio counter-jet reported by \cite{jones86}. 

\begin{figure}
\caption{ Intensity profiles along the extension for each bandpass.
\label{profs}}
\end{figure}
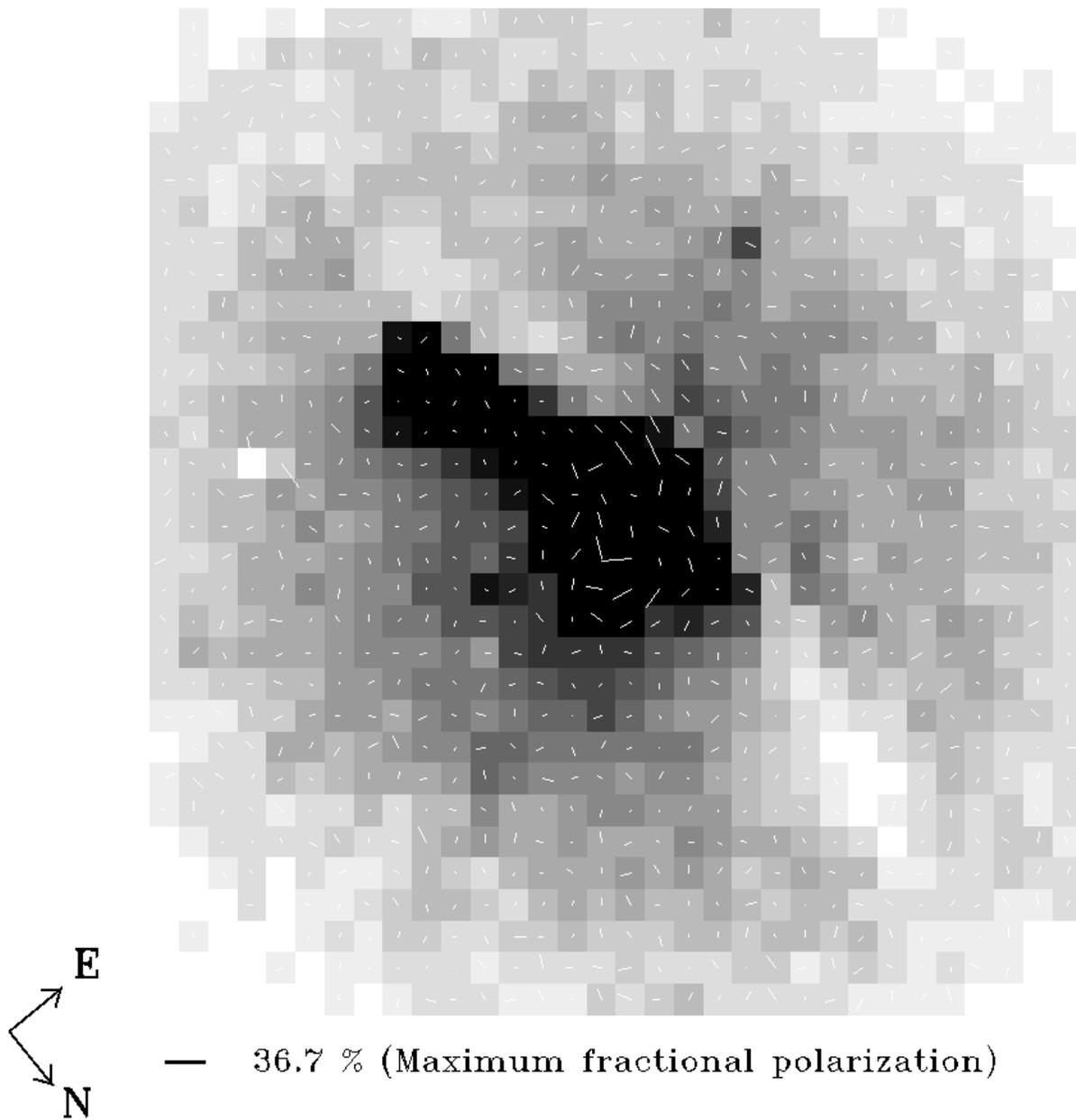
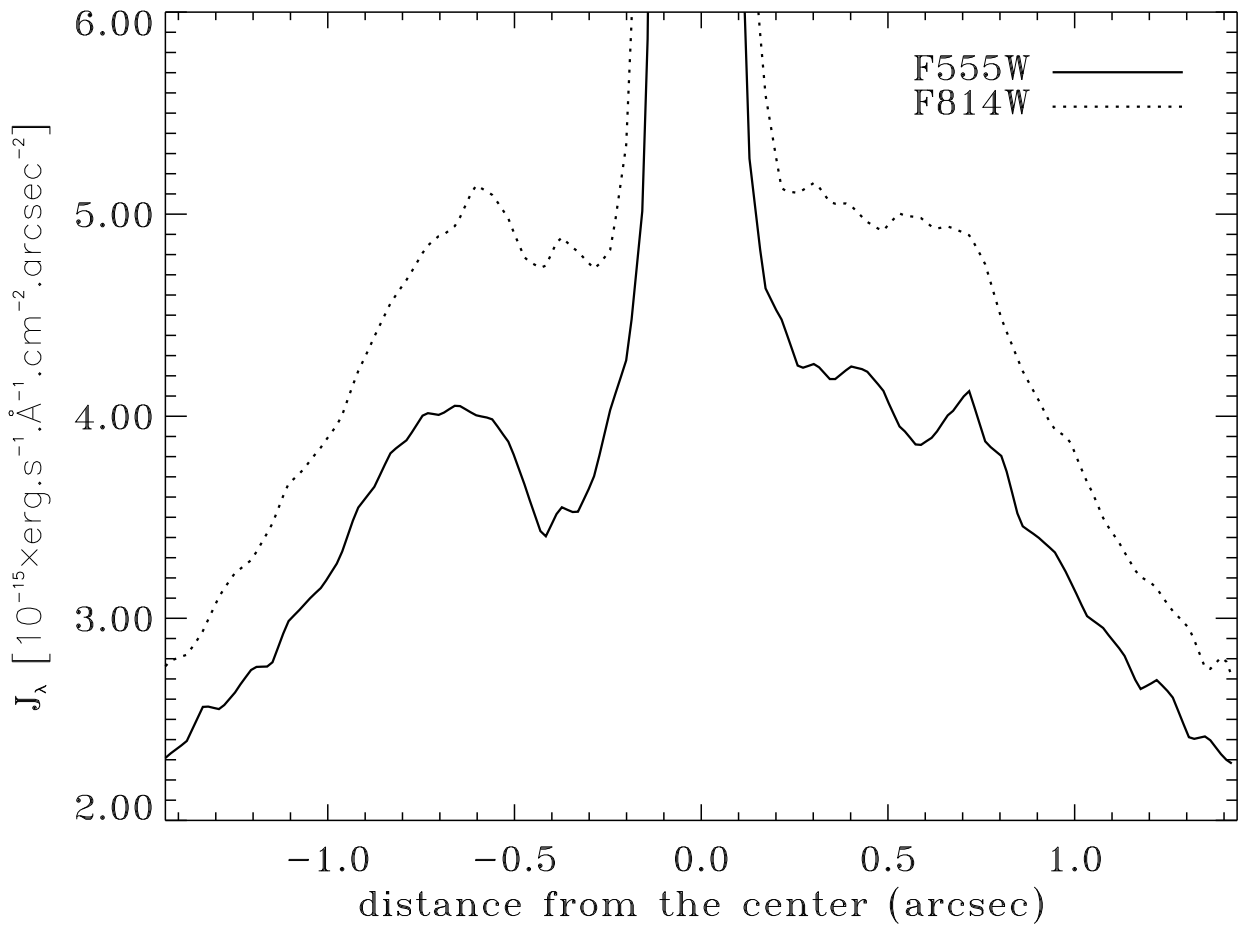
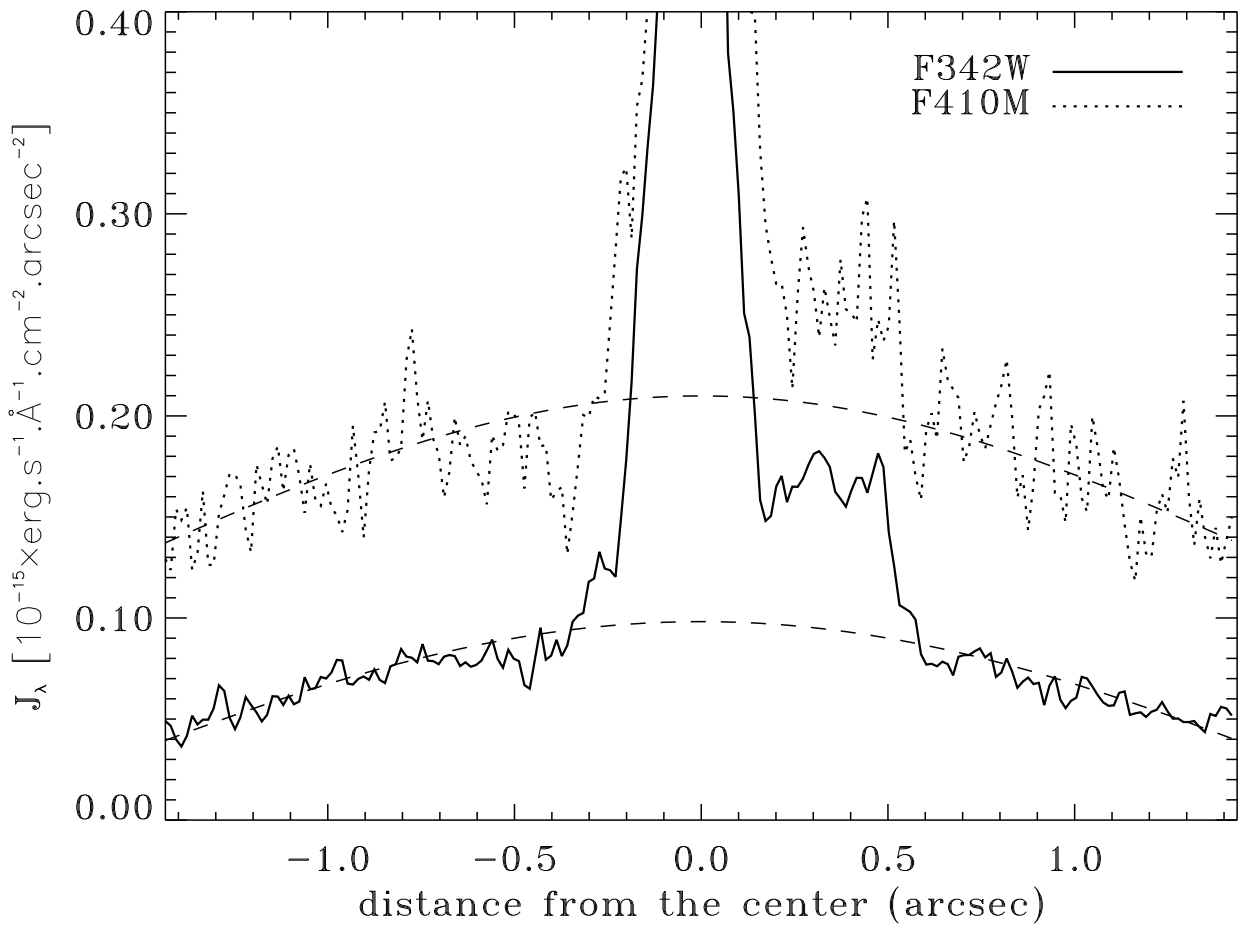

In order to further explore  the origin of this emission, we have performed 
simple photometry by extracting an intensity profile
0.1 arcsec wide along the axis of 
the emission feature in each of the four bands available. 
Figure~\ref{profs} shows these extracted profiles. We note that the observed 
flux is very similar in both the F410M and F342W traces, although the
F410M trace is considerably more noisy. The total fluxes in the extension
are $2.1\pm 0.5$ and
$1.5 \pm 0.5 \times 10^{-18}$ ergs s$^{-1}$ cm$^{-2}$ \AA$^{-1}$  for at
3400\AA\ and 4100\AA\   respectively. These fluxes were determined
in a  0.1 arcsec wide region 
extending from 0.15 to 0.5 arcsec from the nucleus. These may be compared 
to estimated fluxes from the unresolved nucleus of 
${\cal F}(4100) \geq 1.0 \times 10^{-16}$ and 
${\cal F}(3400) \geq 7.3 \times 10^{-17}$ ergs s$^{-1}$ cm$^{-2}$\AA$^{-1}$. 
These are lower limits because we have not 
corrected for the unknown non-linearities in the FOC which may be as large
as 10\%. For the nucleus in the V and R bands, we find fluxes of 
${\cal F}(5550) = 1.0 \times 10^{-16}$ and 
${\cal F}(8140) = 8.6 \times 10^{-17}$ ergs s$^{-1}$ cm$^{-2}$\AA$^{-1}$. 
For the V, or F555 image, this corresponds to $V=18.9$. 

The details of the profiles in Fig.~\ref{profs} led us to question 
whether or not the extension could be seen in the
V band image.  Indeed, a careful inspection of the  V-I color map shows 
some evidence that the extension is also present in the V-band 
image. We determined  an upper limit for the extended emission in the
F555W band of roughly half the flux seen in the F410M and F342W images.

In order to deredden these data, we have used a standard galactic extinction
law. Shuder and Osterbrock(1981) report $E(B-V) = 1.5 \pm 0.6$ determined from 
the Balmer lines which we assume arise very close to the nucleus compared 
to the extended emission. This is an extreme reddenning which we feel is
unlikely to be correct for the extended emission itself, but it is the only
value we have based on actual measurements. We can make an estimate of the 
reddenning by comparison with the Galaxy and assuming that the 
extinction in the  Galaxy and in NGC 6251 are equal.
We determine $A_V = 0.6$ in
the Galaxy from the work of Burstein and Heiles(1981) and 
then assume the same value for NGC6251. With a standard extinction
law, ths corresponds to $E(VB-V) = 0.4$. Without any actual source to 
determine the reddenning, we will use a compromise value of 
$E(B-V) = 1.0 \pm 0.6$ for the following discussion. We emphasize that this 
is a rather uncertain reddening correction. Nevertheless, the basic
conclusions are still valid if we use the lower limit of this reddening
correction. 

Under these assumptions, the
extinction  correction increases the U-band flux relative to the B-band flux
by about 200\%. Thus we find  $U-B = -2.2\pm0.9$. This is indeed a rather
extreme value.

Taking account of the reddening, its uncertainty 
and using the fluxes for F342W and 
F410M given above, we find
the intrinsic spectrum of the extension to be 
$F_\lambda \propto \lambda^{-5.8 \pm 3.1}$ (for unreconstructed
radio astronomers; $F_\nu \propto \nu^{3.8 \pm 3.1}$).  We can use
our upper limit to the flux at $\lambda 5550 $\AA\  to constrain the
exponent in the power law exponent to be $< -1.8$ because otherwise there would
be too much flux at $\lambda 5550$\AA.\   Nevertheless, 
the intrinsic spectrum is likely to be rather blue. 
 
\section{Results}

Several potential explanations for the observations are immediately ruled out
by the absence of radio emission outside the very narrow cone  defined
by the radio jet. In particular, synchrotron emission as seen in the optical 
counterparts to radio jets(\cite{crane93}) is ruled out as well as 
inverse Compton scattering from hot electrons and Thompson 
scattering from an ionized gas. 

We do not favor  emission lines as the origin of the extension. 
First because the bandpass
of the F410M filter contains none of the normal strong emission lines, and yet
the extension is clearly detected in this band( see Fig.~\ref{profs}) with a 
flux close to what is seen in the F342W image. 
Second, the F555W image includes both  the $H\beta$ and  OIII lines. 
Most scenarios for line excitation would produce more flux in the F555W
filter than in the F410M filter. This is not the case so it seems unlikely
that the origin is emission lines.

Another potential origin might be thermal radiation from a population of
extremely hot stars such as cause the UV upturn seen in many galaxies.
The lower limit on $U-B$ color could be appropriate for a collection of 
bright early type stars.
Thus based on the color and total intensity we determine from our 
existing data, 
we cannot rule out the possibility of bright stars
as the origin of the extended emission. However, we feel it is not very 
plausible especially given the implied  weakness of the emission lines.

We consider the most promising explanation to be scattering of radiation 
from a continuum  source off the dust ring itself or off 
material interior to the dust ring. We favor this hypothesis
inspite of the low of polarization in the extended region. 
The high polarization seen close to the
nucleus in Fig.~\ref{poldata} is a major motivation for this 
interpretation. This polarization is an unambiguous indication
that UV light close to the
nucleus  is scattered light. The maximum of the polarized intensity
is concentrated near the radio axis. Although not obvious
in Fig.~\ref{poldata}, there are a few clumps close to the  nucleus that 
show polarizations close to 50\%. Since we believe the radiation both close to
the nucleus and in the extension have  a  similar origin, we  
postulate the observed radiation in the extension is also scattered. The 
observed fractional polarization in the extension is reduced by roughly 
a factor of two from the true fractional polarization by dilution from
the foreground emission in the galaxy itself. Thus the extended region
may have an intrinsic polarization of 10\%.

The main problem with the scattering hypothesis is that the usual
scenarios for radiation emanating from galactic nucleii have the radiation
confined to a cone along the axis of the radio emission. The region
in the plane perpendicular to the radio axis is presumably obscured by
a torus of  material surrounding the active nucleus. In this case it appears
that this obscuration is either not present, or if it is, then we are seeing
radiation that has has been scattered away from the radio axis, and is now
scattered a second time towards us. This double scattering may be the
origin of the extremely blue spectrum. If the scattering hypothesis
turns out to be correct, then these observations do not fit easily into 
standard geometries of active regions.

We should emphasize that none of the proposed scenarios is entirely 
satisfactory, nor are any definitively ruled out except perhaps the 
non-thermal radiation processes which would give rise to radio
emission. This is a particularly frustrating state of affairs.

\section{Discussion}

The canonical models of the nuclear activity in galaxies typically involve 
high energy radiation which is possibly beamed, and interacts with  confining  
material. The geometry and composition of the confining material and 
the viewing angle of the observer are then 
supposed to determine most of the observed properties. 
Observations with the HST have provided us with several good examples which 
tend to confirm this model. Notable among these are the galaxies NGC 1068 
and NGC 4261.

The early HST observations of NGC 4261(\cite{jaf93}) revealed a disk of dust
surrounding the bright nucleus. Subsequent HST observations(\cite{jaf96},
\cite{far96}) have confirmed and
extended these early results. Except for the extended U-band emission, the dust
features seen in the 
images of NGC 6251 are quite similar to those seen in NGC4261. The sizes 
are similar; 240 pc for NGC 4261 and 280 for NGC 6251. The orientations to the 
line of sight are also similar; $64^o$ and $68^o$ respectively. The 
misalignments of the axes of the dust features relative to the radio jet
axes are also similar. These facts suggest a strong similarity in the origin 
of the absorbing dust features seen in these sources. However, this does not 
help to explain the U-band emission seen in NGC6251. The major difference 
between these sources is that the nuclear source on NGC6251 is several 
times brighter than that of NGC4261.

Since the extended emission is very faint and close to a bright
point source, it will be difficult to 
obtain further data that could provide the  clues needed to explain
what we have found here. Clearly  FOC images in other filters 
can provide some of the data we would require. With modern ground based
telescopes,  under excellent seeing conditions, and with 
appropriate spatial sampling, 
it should also be possible to get a reasonable
spectrum of the extension by making use of our knowledge of the spatial
distribution obtained with HST.

Perhaps even more promising than further study of this source, would be 
observations of other promising candidates with the FOC UV filters. Prominent
among these candidates would clearly be NGC4261, but we can easily imagine
several others. One point which we would like to emphasize  in this regard is
the almost entire lack of good high resolution UV images of nearby galaxies 
in the HST program. Our experience has shown that this is an extremely 
rewarding if not perplexing avenue to pursue.

{\bf Acknowledgments} We would like to thank our many colleagues at ESO
and elsewhere who have provided comments on this work. We thank,in particular,
Rick Perley, David Valls-Gabaud,and Hien Tran. 
This work was partially supported by NASA through grant NAS5-2770.

\clearpage
\begin{table*}     
\begin{center}
\caption{ Observational Data}
\begin{tabular}{ccccc}
\tableline
\multicolumn{1}{c}{Camera}    &
\multicolumn{1}{c}{Filters} &
\multicolumn{1}{c}{Exp Time}      &
\multicolumn{1}{c}{Date} &
\multicolumn{1}{c}{Sequence} \\
\tableline
 FOC F/96 & F342W & 1196 & 19 Feb 96& x33t0301 \\
 FOC F/96 & F410M & 1315 & 19 Feb 96 & x33t0302 \\
 FOC F/96 & F342W/POL0    & 2931 &  6 Jun 96 & x37y0301\\
 FOC F/96 & F342W/POL0    & 3222 &  6 Jun 96 & x37y0302\\
 FOC F/96 & F342W/POL60   & 3176 &  6 Jun 96 & x37y0303\\
 FOC F/96 & F342W/POL60   & 3222 &  6 Jun 96 & x37y0304\\
 FOC F/96 & F342W/POL120  & 3176 &  6 Jun 96 & x37y0305\\
 FOC F/96 & F342W/POL120  & 3222 &  6 Jun 96 & x37y0306\\
 PC2      & F555W  &  400 & 28 Jun 95 & u2pq0701\\
 PC2      & F555W  &  400 & 28 Jun 95 & u2pq0703\\
 PC2      & F814W  &  260 & 28 Jun 95 & u2pq0704\\
 PC2      & F814W  &  260 & 28 Jun 95 & u2pq0705\\
 PC2      & F814W  &  200 & 28 Jun 95 & u2pq0706\\
\end{tabular}
\end{center}
\end{table*}

\clearpage
\begin{figure}
\caption{ The upper left panel shows 
the FOC F342W total intensity image obtained from
the polarization data. The upper right panel shows the FOC F401M image, 
while the lower images are the sums of the PC2 F555W(left) 
and F814W(right) images. 
These images have been scaled to the same resolution. 
Each panel covers an area of 2.0 arcsec square. The axis of the radio-jet is
indicated in the upper left panel. It lies 
at a position angle of $300^o$ whereas the axis of the extended emission is 
at  position angle of $\approx 200^o$.
\label{comps}}
\end{figure}
\begin{figure}
\caption{ Greyscale image of the U band image. The dark contours are
determined from the U-band image. The white contours have been selected
from the V-band image to emphasize the dust ring.
\label{dust}}
\end{figure}

\begin{figure}
\caption{ The greyscale shows the total intensity image averaged over
$4\times4$ pixels to give a resolution of 0.056 arcsec.  The arrows show
the direction and magnitude of the polarization. Note that the only regions
of high polarization are symmetrically placed near the nucleus.
\label{poldata}}
\end{figure}
\end{document}